\documentclass[12pt]{article}
\usepackage[utf8]{inputenc}
\usepackage[T1]{fontenc}
\usepackage{graphicx}
\usepackage{caption}
\usepackage{subcaption}
\usepackage[affil-it]{authblk} 
\usepackage[width=150mm,top=25mm,bottom=25mm,bindingoffset=6mm]{geometry}
\usepackage[T1]{fontenc}
\usepackage{amsmath,amssymb,latexsym,marvosym}
\usepackage{pifont}
\usepackage{microtype}
\usepackage{float}
\usepackage{acronym}
\usepackage{multicol}
\usepackage{xcolor}

\usepackage[normalem]{ulem}

\usepackage[style=iso-authoryear]{biblatex}
\begin{document}

\title{\LARGE A democratic Cosmos?}

\author[1]{\large Guilherme Franzmann\normalsize}
\author[2]{\large Yigit Yargic\normalsize}

\affil[1]{\footnotesize Department of Physics, McGill University, Montr\'{e}al, QC, H3A 2T8, Canada}
\affil[2]{\footnotesize Perimeter Institute for Theoretical Physics, 31 Caroline St. N., Waterloo ON, N2L 2Y5, Canada}

\date{}

\maketitle

\begin{abstract}
Despite the success of our best models in Theoretical Physics, especially concerning Cosmology and Particle Physics, we still face persistent challenges. Among them we have the cosmological singularity problem, understanding the late-time acceleration of the Universe, and comprehending the fundamental nature of time. We believe relevant new insights to tackle each of these issues may be found in the Philosophy of Cosmology. We elaborate on three philosophical principles that shall guide us on how to improve our current theories. They are the Copernican Principle for Scales, the Cosmological Heterarchical Principle and the Cosmological Principle of Irreversibility. Following these principles, and using some of our current physical theories as a proxy to implement them, we consider a new assessment of each of these challenges, and show how they may be either explained away, hinting towards new physics, or summarized in a new philosophical principle.
\end{abstract}

\vspace{0.5cm}

\begin{center}
    Essay written for the \emph{2019 Essay Prize in Philosophy of Cosmology\footnote{http://www.rotman.uwo.ca/2019-essay-prize-in-philosophy-of-cosmology/}.} 
\end{center}

\newpage

\tableofcontents 

\vspace{0.5cm}

\subsection*{Acronyms}

\begin{multicols}{2}

\begin{acronym}[SMPP]
\acro{GR}{General Relativity} 
\acro{QM}{Quantum Mechanics}
\acro{SMC}{Standard Model of Cosmology}
\acro{DM}{Dark Matter}
\acro{DE}{Dark Energy}
\acro{CPS}{Copernican Principle for Scales}
\acro{CHP}{Cosmological Heterarchical Principle}
\acro{PCI}{Principle of Cosmological Irreversibility}
\acro{CMB}{Cosmic Microwave Background}
\acro{ST}{String Theory}
\acro{SR}{Structural Realism}
\acro{PRL}{Principle of Relative Locality}
\acro{IR}{Infrared}
\acro{SMPP}{Standard Model of Particle Physics}
\acro{QG}{Quantum Gravity}
\acro{CQ}{Cosmo-Quantum} 
\acro{MOND}{Modified Newtonian Dynamics}

\end{acronym}

\end{multicols}

\newpage

\section[The Cosmic \emph{Pnyka}]{The Cosmic \emph{Pnyka}\footnote{\emph{Pnyka} is a hill in Athens where people would gather around to host their popular democratic assemblies as early as 507 BC.}} \label{sec:intro}
Not so long ago, we used to sit around fires, look at the sky and contemplate the stars, even though we did not yet know what they were. In fact, for a while, it was even thought that the stars were just other people sitting around their own fires. Today, apart from our technological advancements, we still remain sitting on \textit{our} planet Earth, around \textit{our} fiery Sun, and pondering about \textit{our} place in the Cosmos.

Fortunately, though, the tale of the Universe has kept evolving since that time. We have come a long way in our understanding of the inner workings of the Cosmos, by making use of both metaphysical arguments as well as scientific observations and models. Still, we should remember that the latter endeavor is a much more recent attempt to understanding the Universe, and that it has only predominantly taken over that role in the last few decades.

Cosmology, as we call the field of study concerned with the inquiry of the Cosmos, has become a successful scientific field only in the 20th century. Concerned with the origin, evolution and structure of the Universe, it comes as no surprise that it developed this late in human history given the technological requirements needed for the study of the fossils in the sky. 

Considered a branch of Physics, Cosmology has its own peculiarities. It is not by chance that we have already referred to our quest as a \emph{tale}, to some insights as being of \emph{metaphysical} nature, and to our observations as \emph{fossils}. That is just another way of saying that Cosmology shares its paradigm with other fields of knowledge.

The singular existence of the Universe\footnote{We take a conservative stance regarding the multiverse which is aligned with \parencite{edgeEllis2017}. However, we will briefly comment on the possible existence of the multiverse in Section \ref{sec:Copernican}.} \parencite{unger2015singular} together with its absolute inclusion of everything that exists distinguish it from any other physical system that can be studied in the laboratory. Our incapacity for controlled intervention and ensemble observation of such a system challenges most of the successful methods in Physics, and the whole scientific framework more broadly. The typical laboratory conditions which are considered for experiments in most of the natural sciences are untenable, such that naive empiricism drops out of the picture completely. In fact, it is still an open debate whether Cosmology is, therefore, underdetermined\footnote{The underdetermination of a scientific theory by evidence is the idea that the available evidence at a given time may not be sufficient to determine the set of beliefs one should hold in response to it \parencite{sep-scientific-underdetermination}.} \parencite{sep-scientific-underdetermination,sep-cosmology} due to these inherent difficulties. On the other hand, this presents an opportunity for other inputs to be considered towards this endeavor. Indeed, the very fact that there is a single Universe together with the finiteness of the speed of light makes Cosmology an archaeological, historical science (also quite similar to evolutionary biology\footnote{See for instance \parencite{Smolin:1994vb}, where a cosmological scenario is discussed for which the parameters of the Standard Model, both particle and cosmological, are set due to a fitness problem that extremizes the production of black holes.}), while the lack of experimental intervention calls for simulations and further philosophical propositions. 

Outsourcing Theoretical Physics, and in particular Cosmology, to philosophical principles is not anything new. One of the best examples of such a practice dates back to Einstein. Part of his ingenuity in formulating General Relativity (GR) resided in his ability to consider philosophical assertions about nature and let them guide the development of his theory. These assertions were borne out of Ernest Mach's criticism of the notions of space and time in Newtonian mechanics. Notice that, although GR is empirically accessible nowadays, it was not at his time. Thus, it is quite encouraging to know how far philosophical considerations have taken us and helped us navigate the landscape of new theories of nature in the past.

Yet, somehow the above discussion seems to fall short for a large part of the theoretical physics community\footnote{See for instance Steven Weinberg's book \emph{Dreams of a final theory }\parencite{Weinberg94}, where Chapter VIII's title is \emph{Against Philosophy.}}. Reminders such as ``\emph{Physics needs philosophy, philosophy needs physics}'' \parencite{2018FoPh...48..481R} and ``\emph{Theoretical physics and cosmology find themselves in a strange place}'' \parencite{edgeEllis2017} tells us that something peculiar is going on. Carlo Rovelli's message may have seemed obvious in the early 20th century, but it is no longer the case today. As our most modern theories, namely Quantum Mechanics (QM) and Relativity, kept evolving through new generations of researchers, it seems that the community has slowly departed from philosophical circles, and has changed the balance of their sources of inspiration to trusting mathematical consistency more and philosophical arguments less.

This perspective may be well-justified. After all, the Standard Model of Cosmology (SMC) is quite successful. Relying solely on six free parameters, it is able to account for most of the current data, which has become abundant for the last 30 years starting with the COBE (Cosmic Background Explorer) mission. It also provides an exquisite description of the evolution of the Universe that extends from many fractions of a second to its current age, around $13.8$ billion years \parencite{Ade:2015xua}. It is noteworthy that the Universe's age is one of those free parameters in the SMC.

Paradoxically, the success of the SMC is also its Achilles' heel. Its potential to provide us with a coherent picture of the Universe relies on mysterious forms of matter and energy, Dark Matter (DM) and Dark Energy (DE), which remain elusive to our understanding, despite being responsible for around $95\%$ of the Universe's energy budget. Moreover, the resulting dynamics of such components seems puzzling by itself, given we happen to live in a period of our Universe's long history when only recently both DM and DE were coincidentally just as relevant for its evolution. This is usually referred as the Coincidence Problem (see for instance \cite{ArkaniHamed:2000tc}). There are at least three other problems we also need to face. They are the singularity problem \parencite{Hawking:1969sw}, clarifying the fundamental nature of time \parencite{Penrose:1980ge,Price:1996}, and the nature of DM and DE.

To some of these problems we intend to offer insight. That is not an easy task, since we aim to do so in a concise way, relying on a few philosophical principles. These principles are the Copernican Principle for Scales (CPS), the Cosmological Heterarchical Principle (CHP), and the Principle of Cosmological Irreversibility (PCI). Once we have elucidated them, we will use proxy theories as means of their implementation, and show how they can possibly alleviate some of the problems mentioned above.

\begin{center}
\ding{102}
\end{center}

In many ways, the Cosmos is a Pnyka. Not only does it incite us to bring together different branches of Physics, but also it motivates us to consider several fields of knowledge, and their respective paradigms, should we hope to understand it. As democratic as such an approach might look, we will argue that our current account of it is not as fulfilling as we would wish, and that may be preventing us from achieving new insights to overcome the current problems. By imposing a more egalitarian account of the Cosmos, we will argue that a much more coherent and maybe even predictable cosmological scenario might be around the corner.

\section{The Copernican Principle for Scales}\label{sec:Copernican}
One of the most haunting questions that we have ever asked is ``Where do we come from?''. To a large extent, this inquiry has motivated people to all sorts of endeavors, from telling diverse myths of creation to waging wars against one another.

Historically, one way of addressing this question is seeking to expand the perspective of our place in the Cosmos. A few representations of this trend can be roughly laid out as follows:  
\begin{itemize}
    \item[1.] First we considered the Earth to be at the center of the Cosmos (Ptolemaic system);
    
    \item[2.] Then we assumed the Sun to be at the center of the Cosmos (Copernican system);
    
    \item[3.] Later we learned the Solar System was just one among many within our Milky Way galaxy after Galileo Galilei's observations with the telescope (Galilean System);
    
    \item[4.] Less than a hundred years ago, we realized there were many galaxies, and ours was not special. We discovered the Universe (Kant-Hubble system\footnote{Immanuel Kant already suspected in 1755 in his \emph{Universal Natural History and Theory of the Heavens} that the Andromeda Galaxy, considered a nebulae at his time, was external to the Milky Way. Only after Edwin Hubble's observations published in 1929 we finally learned that, indeed, the Universe is made up of many more galaxies like our own.}). 
    
\end{itemize}
Each of these steps is a representation of what we call the Copernican Principle, which is an alternative name to the principle of relativity, in the sense that privileged, sometimes absolute structures are promoted to become relational ones. Note that our formulation of the Copernican Principle is slightly different than the idea that we, humans, are not privileged observers in the Universe. We consider the latter to be a corollary of the former.

Not surprisingly, each of these stages was accompanied by specific ways of understanding the world which were also being ``Copernicized'':

\begin{itemize}
    \item[i.] The laws on Earth and elsewhere were not considered to be the same. In particular, Earth was in absolute rest while everything else was moving in the celestial spheres. Time and space are complete absolutes;
    
    \item[ii-iii.] It is learned that it is the Earth which is moving, together with the other celestial bodies in the Solar System, around the Sun. And all of them follow the same laws, which are the same found to be followed by us on Earth. Thus, our mundane laws are not special; time and space remain absolutes;
    
    \item[iv.] We discovered that time and space are not absolutes, in fact they are better understood as intertwined into a unique structure called spacetime which is \emph{determined} by the very existing matter inhabiting it\footnote{A more rigorous ontological account of the differences between each of these scenarios can be found in \parencite{Esfeld2008-ESFMSR}.}. 
\end{itemize}
The attempt to ``Copernicize'' our physical descriptions of nature is what we call nowadays the Principle of Increased Background Independence, which is the attempt to eliminate fixed and unchanged structures in favor of describing relationships between dynamical entities \parencite{Smolin:2005mq}. It begs the question how each of the above lists shall evolve, and how each new item is related to the profound problems we encounter currently in Theoretical Physics and Cosmology.

One possible answer, which many physicists see as a next step towards Copernicizing our understanding of the world, is to promote our Universe to be just one among infinitely many. The idea of the multiverse closely follows one theory which may account for many features of our current Universe, the theory of inflation\footnote{Inflation summons an accelerated expansion phase in the early Universe that explains why the Universe we live in seems to be so spatially flat, so large and how the fluctuations in the Cosmic Microwave Background (CMB) are almost scale-invariant, therefore also explaining how structures have been formed in our Universe. Most of the inflationary models rely in a quasi-de Sitter initial phase that smoothly shifts towards standard radiation decelerated expansion.} \parencite{Starobinsky:1980te,Guth:1980zm,Linde:1981mu,Steinhardt:1982kg,Vilenkin:1983xq}, though it also finds a basis of existence in the vacua landscape of String Theory (ST) \parencite{Susskind:2003kw}. An older version of the multiverse idea with quite different flavour goes back to Everett's interpretation of QM \parencite{Everett1973-EVETTO-2}.

Our account here is radically different, and it is aligned with the philosophy that has been tacitly assumed over the course of scientific history. The next step towards a clear understanding of our place in the Cosmos is not extrapolating known theories to their ultimate consequences, which is to a large extent what motivated the introduction of the different realizations of the multiverse, but rather continuing to substitute absolute structures within those theories for more relational ones. How can we do that? 

As a starting point, it seems reasonable to say that we do not have access to the intrinsic nature of physical entities, therefore we do not have access to the intrinsic nature of the World. Moreover, it seems that we only have access to the relational properties of physical entities \parencite{sep-structural-realism}. This philosophical position is usually referred to as Structural Realism (SR) and it comes in many shapes. Among them, Epistemological Structural Realism states that this is simply an epistemic condition, while Ontic Structural Realism emphasizes the ontological priority of relations over entities.  

The theory of Relativity can be seen as an implementation of such a program, to the extent that its observables become relative to their observers, instead of remaining intrinsically pristine regardless of the conditions under which they are observed. The full program of SR can be seen as a guide to our understanding of the world becoming increasingly more relational, so that physical theories rely less on absolute structures. 

With that in mind, we may start considering to lay down the next items in our lists above. The first thing we bring to attention is the empirical fact that we do not ever measure space, since all the  events we observe are actually not directly observed as being separated from our measurement devices. As \parencite{AmelinoCamelia:2011bm} phrases it, we are \emph{calorimeters with clocks}, and all that we fundamentally measure are energies and angles of the quanta we emit and absorb, as well as the time of those events. This kind of assessment led to the development of the Principle of Relative Locality (PRL) (Ibid.). 

The idea can be summarized after remembering how we reconstruct spacetime coordinates using light signals in Relativity: we consider the time photons take to travel back and forth between events, but we disregard information about their energy. This tacitly implies that the same spacetime would be reconstructed regardless of the frequency of the light signals being used. This simple assumption results in having momentum space being a linear manifold, which is shown to be related to a notion of absolute locality, \emph{i.e.}, for which all different observes agree with their spacetime reconstruction. Thus, when this assumption is dropped and the PRL is implemented, the geometry of momentum space becomes richer, curvature is introduced, and different observers no longer agree with their spacetime reconstruction, resulting in locality violations. 

The PRL opens precedence for the Copernican Principle for Scales. We note that our current theories of Nature successfully account for all phenomena in a wide range of length and energy scales in which they have been tested. While the Large Hadron Collider probes subatomic particles down to $10^{-19}$ meters, the Hubble telescope can observe structures up to $10^{26}$ meters. Our size is roughly at the center of this range, likely not because humans are special, but because this is our starting point. 

Typically, when we think about scales we tend to think of a semi-straight line. Take length scale for example. Regardless of whether we live in a continuum manifold or having an underlying discrete grid where physical objects are defined (usually believed to be of the size of the Planck length), both scenarios imply a lower boundary for the smallest conceivable length, while the largest length remains unbounded. Therefore, when we plot this over the real line, we observe that our typical length scale of one meter\footnote{The argument does not depend on the units being considered, of course.} can be seen as special in the sense that it is infinitely closer to the lower bound than to infinity. 

One way of avoiding the arbitrariness of choosing our scale to be special is to consider a logarithmic scale instead. We implement this idea in Section \ref{sec:landscape}. By doing so, we can make more symmetric considerations about the small/ultraviolet and the large/infrared, for instance.  In particular, it becomes quite clear that if we impose a Copernican principle for the length scale, we should not expect our scale to be special in the sense that it emerges as a divide for which new descriptions of nature are necessary. So far, we observe that we have had major developments of new theories towards the ultraviolet, while the infrared (IR) has remained mostly untouched. 

Therefore, we propose \textbf{the Copernican Principle for Scales}: \emph{The human's scale is not special. However, since our devices are built having our scale as a starting point, defining it to be a pivot scale, then we should expect new physical theories falling on both sides of this arbitrary division}.

At the moment, we seem to be in violation of the CPS, since most of our physical theories have sought completion of ultraviolet physics. We will argue in Section \ref{sec:lambda} that our current observations, taken in light of CPS, may be telling us that a new theory may be at our door in the deep infrared.

\section{The Cosmological Heterarchical Principle}\label{sec:heterarchy}
Although the world could seemingly be described either hierarchically or heterarchically, human's reasoning and intuition brought us an account of the world for which hierarchies play a major role. It comes without surprise that we can find many hierarchical descriptions of the world throughout human history. For instance, \emph{The Great Chain of Being}, having God at the top and going all the way down to minerals, was considered in medieval Christianity; another example can be found in Leibniz's monads \parencite{sep-leibniz}, considered to be the building blocks of the Universe, which were ordered as created monads, monads with perception and memory (souls), rational souls (spirits), and finally the omniscient monad, God. 

Once we start considering a hierarchical framework, a notion of reducibility follows almost immediately, where we expect that higher structures could, in principle, be reduced to lower ones. The bridges connecting these layers have a variety of names in different fields of knowledge; in Physics these include derivation, emergence, coarse graining, reduction, among others. Note that those names may also indicate whether we are climbing up or down the hierarchy's stairs. 

Following this analogy, it seems natural that our attention is drawn to at least four questions: i) Do the stairs have a bottom? ii) Do the stairs have a top? iii) What are the qualitative properties, and corresponding quantitative increase/decrease, as different steps are considered? iv) Is there a pivot step\footnote{Note that this question is a reflection upon a generalized Copernican principle.}? It is not clear if the answers to any of those questions are beyond practical examples, being invariant if we consider different fields of knowledge and, therefore, reflecting on a general ontology of the world. Although we will be focusing on Physics, and in particular Cosmology, we may try to at least lay down some general possible scenarios.


Let us start with the easiest question, iii). The answer is subjected to the field under consideration. For instance, in late-time Cosmology we have astrophysical objects, like stars, which are brought together gravitationally to form galaxies; those galaxies can also be brought together by gravity to form cluster of galaxies; and even those clusters can agglomerate to form superclusters of galaxies, the largest known structures to date. Thus, for this particular case, each step up in the hierarchy is defined by a change in the center of mass, which establishes the orbit (qualitative property) for those larger objects (quantitative increment). 

After our discussion of the Copernican Principle in Section \ref{sec:Copernican}, question iv) also becomes easier to tackle. It seems that a reference step in a given hierarchy will always emerge given that we experience the world from a very particular condition of existence, far removed from an absolute, transcendental existence. The challenge is, therefore, to make that explicit and to be able to overcome such a privileged starting point in order to make progress towards understanding the remaining steps. Notice that the very presence of a hierarchy implies that steps up/down a given reference step will not be symmetrically understood. 

Finally, the first and second questions, whether the stairs of hierarchies are somehow bounded, are the hardest questions. It seems that the answer is either yes or no. These potentially obvious answers have non-trivial implications for their respective subscribers and their consequent approach to formulate theories. If the answer is no, then we have a framework subjected to an infinite regress\footnote{From \parencite{sep-infinite-regress}: ``An infinite regress is a series of appropriately related elements with a first member but no last member, where each element leads to or generates the next in some sense. An infinite regress argument is an argument that makes appeal to an infinite regress. Usually such arguments take the form of objections to a theory, with the fact that the theory implies an infinite regress being taken to be objectionable''. Note that the direction the regress goes in is not relevant. }. On the other hand, if the answer is yes, then there is at least one bottom/top step which sustains itself on its own, referring all possible inquires about further down/up steps back to themselves (unavoidably introducing self-reference), and somehow defining a sort of ``\emph{bootstrapping ontology/epistemology}''\footnote{Note that we use  \emph{quotes} and \emph{slash} because we are not being rigorous in defining precisely which aspects correspond to either the assumed ontology of the framework or to its epistemology. In fact, from a SR point of view, those two things are naively overlapping.}.

This last notion is equivalent to a heterarchy. We will not get into the details of how that works in general (though it is discussed thoroughly in \parencite{Hofstadter:1979:GEB:539932} in the context of Philosophy of Mind and Computer Sciences), but looking at an example can help. The very notion of a particle in the Standard Model of Particle Physics (SMPP) after renormalization implies that no particle can be defined without referring to all other particles, since virtual particles, following well defined interaction rules provided by Feynman diagrams, can pop up arbitrarily and recursively in any physical process. Thus, the bottom step of the hierarchy present in Particle Physics is composed of many fundamental particles having many recursive loops. Note that higher levels in this hierarchy can be completely sealed off of such tangling. Another prominent example can be found in the context of Molecular Biology in the Central Dogma (see Ibid, pg. 533).

The underlying message is that we need to seek theories which avoid infinite regress by making use of heterarchical structures. This is already the case in Quantum Field Theory, but should also be the case for a Quantum Gravity (QG) theory and a Cosmological theory. 

Therefore, we propose \textbf{the Cosmological Heterarchical Principle:} \emph{A Cosmological theory should not allow any form of infinite regress, thus it should be hierarchically bounded on both ends.}

The implication of such a principle will be explored in Section \ref{sec:t_duality}, where we will see that ST seems to provide the necessary ingredients to establish a hierarchically lower bounded theory of spacetime by making use of T-duality.

\section{The landscape of scales}\label{sec:landscape}
As intelligent beings, we interact with our surroundings and observe the World we live in. These interactions accumulate with our experience about where we are and how this place works. Our minds work consciously and subconsciously to deduce an underlying structure in which we understand the World. Yet our understanding of the World is inherently attached to and limited by what we are able to observe.

According to the best of our knowledge today, we live on a floating rock in space roughly 14 billion years after a very hot and dense period of the Universe. This empirical knowledge was not available to most of our ancestors, who were not able to observe the things that we do now: the things that are further, larger, smaller, faster... We expanded our abilities to observe phenomena at a wide range of scales and reached our current understanding.

As we stated before, we are able to describe phenomena spanning a range of 45 orders of magnitude, from subatomic particles to the deep outer Cosmos. This range of scales is our territory of understanding. We have the greatest control around its center, where we are located, and our control diminishes near its borders. Yet we want to explore and expand our understanding even further.

As we explore new territories in the landscape of scales, we frequently discover new phenomena that upgrade our knowledge. We recognize a discovery when the observation starts to deviate from our previous knowledge.

Two of the greatest revolutions in modern physics are attached to the discovery of boundaries of the landscape of scales. The first one is the speed of light, $c$, which is the upper bound for the speed of any object. The second one is the Planck constant, $\hbar$, which sets the limit to the precision by which the position and momentum of an elementary object can be measured simultaneously. These are two fundamental constants of the World we live in and which set its boundaries.

There are likely at least two more fundamental constants, which come from GR, and become relevant on larger scales: the gravitational constant $G$ and the cosmological constant $\Lambda$. These two can also be understood as setting boundaries to the landscape of scales: the energy inside a spherical volume per its radius cannot exceed $c^4/(2G)$, at which point a black hole is formed, marking the limit for maximum compression; similarly, the total energy density inside a volume cannot be made lower than $\rho_\Lambda = \frac{c^4 \Lambda}{8\pi G}$, at which point the only contribution comes from the cosmological constant\footnote{The cosmological constant $\Lambda$ also sets the radius of the de Sitter horizon $\sqrt{3/\Lambda}$, but we preferred the density interpretation here.}.

\begin{figure}[H]
    \centering
    \includegraphics[width = \textwidth]{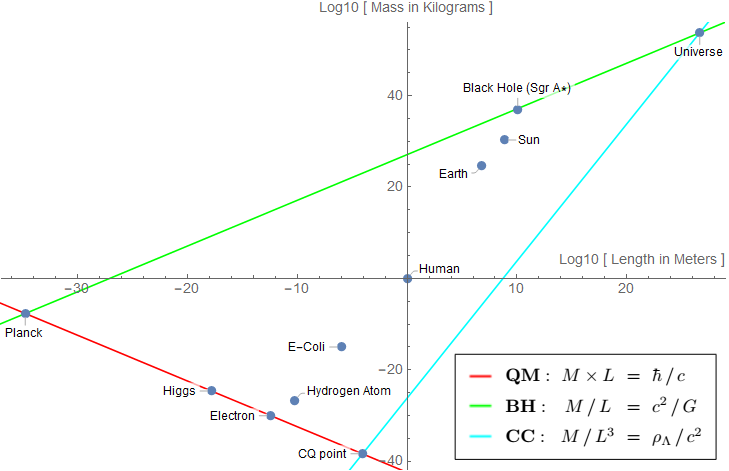}
    \caption{Characteristic length and mass scales for various objects. The region is surrounded by three boundary lines coming from Quantum Mechanics, Black Hole Physics and Cosmology.}
    \label{fig:Scales}
\end{figure}
We illustrate the aforementioned ideas in Figure \ref{fig:Scales}, in which the characteristic sizes and masses of various physical objects are plotted on a logarithmic scale, following the discussions contained in Section \ref{sec:Copernican}. The examples include humans at the center (who adjust the scales according to themselves), astronomical objects (Earth, Sun and black hole), microscopic objects (bacteria and atoms) and elementary particles (electron and the Higgs boson).

There are three lines in Figure \ref{fig:Scales}. The red line marks the quantum mechanical boundary $\mathrm{mass} \times \mathrm{length} = \hbar/c$. The elementary particles sit here. The green line marks the black hole boundary $\mathrm{mass} / \mathrm{length} = c^2/G$. Finally, the light blue line marks the boundary $\mathrm{mass}/\mathrm{length}^3 = \rho_\Lambda / c^2$ from the corresponding cosmological constant density. 

The intersection points of these lines are of great physical importance. The observable Universe, which is today in its quasi-de Sitter phase, lies roughly where the black hole and cosmological constant lines meet. The black hole and quantum lines meet at the Planck scale, where QG is expected to reign.

In contrast, there is only very little discussion in the literature around the third intersection point, where the cosmological constant and quantum lines meet\footnote{Though see \parencite{Smolin:2017kkb}.}. This scale is given by the constants $\hbar$, $c$ and $\rho_\Lambda$ and it deserves a name. We will call it here the ``CQ point'', which is an acronym for ``cosmological'' and ``quantum''.

The lightest massive elementary particles in the Standard Model are the neutrinos. We know the neutrinos have a mass below a few hundred mili-electronvolts, but its value has so far been beyond experimental reach. The CQ energy $(\hbar^3 c^3 \rho_\Lambda)^{1/4} = 2.3 \, \mathrm{meV}$ lies only two orders of magnitude below the current upper bounds for neutrino masses. Hence, this is unexplored territory for elementary particle physics.

\begin{center}
\ding{98} 
\end{center}

Having introduced the philosophical principles above, and 
laid out explicitly the landscape of scales in this section summarized in Figure \ref{fig:Scales}, we are finally able to start discussing some of the current problems in Cosmology. We will start by looking at the problem of the singularity in the very early Universe and then turn our eyes to late-time Cosmology, in particular to the intriguing intergalactic MOND-like dynamics and its relation to the current accelerated expansion. 

\section{The Very Early Universe}\label{sec:t_duality}
One of the problems of SMC is that it is singular. Despite its success, its reliance on GR makes this problem unavoidable, so it is not a surprise that considering inflationary cosmology for the early universe has not explained this issue away \parencite{Borde:1993xh,Borde:2001nh}. Currently, it is largely expected that only a fully-fledged theory of QG should yield a non-singular cosmology. 

The singularity problem can be interpreted as a failed attempt to tackle a temporal infinite regress, which among other consequences encodes the idea that every effect is preceded by a cause, continuing \emph{ad infinitum}. As we have seen in Section \ref{sec:heterarchy}, the only way to handle an infinite regress in a given hierarchy, in this case temporal evolution, is by imposing some hierarchical bounded structure which is self-sustainable. The known examples always imply some tangling between a few levels of a given hierarchy through self-referential relations. Thus, this is the kind of structure we should look for while trying to solve the cosmological singularity problem. 

It turns out that we have a proxy for such framework in ST, through the notion of T-duality for closed strings. In fact, we will see that it also incorporates what we have learned about the CPS, since it implements a notion of the PRL. Let us have a look at how that works. 

\subsection{T-duality}

To keep it simple, let us consider only the bosonic closed sector of the theory, for which we know that the mass spectrum of different string excitations for a string wound around a compact dimension, taken to be a circle of radius R, is given by, 

\begin{equation}
    M^2 = (N + \Tilde{N} - 2) + \frac{p^2}{R^2} + \frac{w^2 R^2}{\alpha'^2}\nonumber.
\end{equation}
The first term accounts for the oscillatory modes of the string, which can propagate counter- and clockwise, the second term accounts for the momentum modes, denoted by $p$, which are the same as if we had considered a wave inside a box of side length $R$. The third term is the novelty, since it can only appear due to the fact that the string can wind around the compact dimension, resulting in a topological charge called winding number, $w$, so that we typically say this contribution is due to the winding modes. The constant $\alpha'$, proportional to the inverse of the string tension, characterizes the length scale of the strings, $l_s \sim \sqrt{\alpha'}$, which is larger (usually by some orders of magnitude, \cite{Baumann:2014nda}) than the Planck length, $l_P$, considered to be the smallest possible length in the framework of QG. 

The surprising aspect of the presence of the winding modes is that its energy contribution is proportional to the radius of the compact dimension, in contrast to the momentum modes' energy, which is inversely proportional to the radius. Thus, it is reasonable to expect that as the box expands and shrinks, different modes are excited, which can be easily understood by the simple assumption that nature tends to privilege energetically favorable configurations.

The duality appears when we look to the mass spectrum and realize that it remains invariant under, 
\begin{align} \label{eq:duality}
\begin{split}
    p & \longleftrightarrow  w \\
    R & \longleftrightarrow  \alpha'/R
    ,
\end{split}
\end{align}
which exchanges the number of momentum and winding modes as well as consider the reciprocal of the radius to be scaled by $\alpha'$. Although we show this duality here for a very simple example, it is actually a duality of the full theory, including any physical process under consideration.

The most important consequence of such mode-availability is that the notion of position becomes a derived concept. An easy way to understand that is to imagine that in order to measure distances, for instance, we need to build a photon wave-package, which depends on the Fourier modes available. However, as we have just seen, those are dependent on the size of the box in which the observer finds herself. This has been already pointed out in the seminal work on String Gas Cosmology \parencite{Brandenberger:1988aj}, where the notion of two position operators was introduced, 

\begin{align}
    |x\rangle &= \sum_p e^{i x \cdot p} |p\rangle \nonumber\\
    |\tilde{x}\rangle &= \sum_w e^{i \tilde{x} \cdot w} |w\rangle, \nonumber
\end{align}
while the physical position operator, $ |x_p\rangle$, would correspond to a linear combination of them depending on the size of the box. In fact, nowadays this dual coordinate notion has also been explored in cosmological backgrounds for the time coordinate \parencite{Brandenberger:2018bdc,Bernardo:2019pnq}.

The progress we make relying on T-duality is in our account of what spacetime really is and how it can be reconstructed. As it turns out, the Fourier transform of momentum space, historically referred to as target space, to which we usually attribute the notion of the physical spacetime is shaken, for now there is an analogous transformation we can consider for the winding modes, defining a \emph{winding space}. 

Thus, there are two notions of space: \emph{winding space} and \emph{target space}. The duality between them given the above transformation makes the definition of the physical spacetime subtler \parencite{Huggett:2015vla}. The reason for that is that when an observer probes the World, she will do it using the modes which are available to her: when momentum modes are available, target space will match physical space; while when she uses winding modes for the reconstruction, then winding space will match the physical space.

\begin{figure}
    \centering
    \includegraphics[scale=0.6]{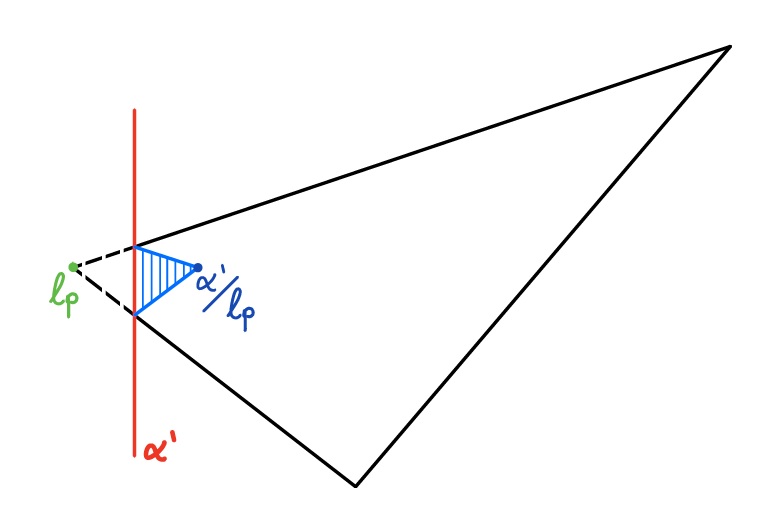}
    \caption{A schematic representation of Figure \ref{fig:Scales} suitable for our interest in the early universe. The triangular region to the left of the pivot length scale defined by $\alpha'$ up to the minimum Planck length, $l_P$, is mirrored to the right side of such scale as the blue triangle. The already known dynamics of that region would account for the dynamics of the blue triangle, therefore providing the dynamics of the region between $l_P$ and $\alpha'$.}
    \label{fig:alphaprime}
\end{figure}

The key point is to realize that winding and target spaces are dual to each other by (\ref{eq:duality}). Hence, even though for small $R<l_s$ momentum modes are not available and the reconstruction is made using winding modes with an effective radius of $\alpha'/R$, this does correspond to what would be the dynamics of having momentum modes in a space with radius $R$. Thus, the seemingly inaccessible region $l_s > R > l_P$ in target space can be reconstructed by using winding space with an effective radius $l_s< \tilde{R} < \alpha'/l_P$. That is schematically showed in Figure \ref{fig:alphaprime}.

This intuitive picture provides us with the necessary ingredients of a ``bootstrapping ontology/epistemology'', since the bottom steps of the hierarchy are tangled as expected after the discussion on Section \ref{sec:heterarchy}. Something similar should be expected for the other vertices of the triangle of scales. 

As it turns out, T-duality does imply that the cosmological singularity is removed. We will not discuss how that is the case here, since we were mostly concerned in making explicit its ontological/epistemological grounds. The reader interested in seeing how that is the case can find a recent review in \parencite{Brandenberger:2018wbg} and references therein. 

\section{{\texorpdfstring{$\Lambda$}{L}}-theory: GR's unknown heir in the deep IR} \label{sec:lambda}

Finally, we argue for the possibility of a paradigm shift in Cosmology coming from theoretical developments in the deep infrared regime. The motivation comes from philosophical arguments, especially the CPS, as well as several puzzles in modern physics including DE and DM, which may need a new framework for their resolution.

GR dominates our modern understanding of the spacetime geometry and gravitational physics. It is often celebrated as one of the most mathematically elegant and empirically successful theories. However, despite its achievements, the realm of GR is still an active battleground for many fundamental questions. These questions provide the motivation for looking for a more fundamental theory that shall replace GR. Most effort in this direction is aimed at its ultraviolet completion due to its infamous discrepancy with QM, as well as the black hole and early universe singularities, where GR is expected to break down. Here, we will do the opposite and question the validity of GR on very large scales.

As we mentioned in the Section \ref{sec:Copernican}, the study of Cosmos is subjected to a different scientific status than the rest of Physics due to its singular existence. The theories that are able to contain the Universe as a solution, GR in particular, cannot be tested directly with varying inputs as in a typical laboratory experiment or astronomical observation. Instead, we can test its validity only on astronomical objects and extrapolate the theory to the cosmic scales.

This does not mean that nothing can be observed about the Cosmos. Edwin Hubble's discovery of the cosmic expansion as well as the study of the early universe are famous examples of how we infer knowledge and test our theories about the Cosmos indirectly.

The real problem is that the cosmic puzzles are often difficult to separate from the rest of Physics due to its inclusion of everything.
This allows room for patching the small-scale physics without modifying the theories of the large. This has been the case at least twice: for the problems that led to the development of inflationary cosmology, and for the problems that led to the DM hypothesis.

Both inflation and DM are successful in being able to explain various observations, therefore they are widely accepted among the physics community. However, their success does not guarantee that they are the right explanation for those phenomena, since both theories are largely underdetermined by data and their origins remain unknown. All the evidence for them are read and interpreted with the lenses of GR, hence they rely on the validity of GR on large scales.

GR has been widely tested on intermediate scales, but not as much on cosmic ones. If GR is supposed to be replaced on the large scales, we need to find observations that parametrically deviate from its predictions. Do we observe any deviations from GR at this scale yet?


The answer to this question depends on understanding the dark components (DE and DM) of the Universe. Although SMC keeps GR intact while addressing these problems, it is also natural to suspect the validity of GR at the cosmological scale and expect a more fundamental resolution. Our hope is that an extension of GR at the cosmological scale, dubbed ``$\Lambda$-theory'', will connect DE and DM to each other, or to a common source.

There is nothing new here. Many extensions of GR have been proposed that would account for different dynamics at the large, and still recovering proper observational limits that just recovers GR. This underdetermination of the correct extension is due to, among other factors, the enormous difference in scales between astrophysical observations and the scale of the Universe. In order to have a better guidance in exploring the depths of the IR and possibly defining a north in landscape of theories, we bring attention to a class of phenomena that remains puzzling in astrophysical scales that seems to be coupled to cosmological scales. And for that, we take MOdified Newtonian Dynamics (MOND) as a proxy for their description.

\subsection{Inspirations from MOND}

Why should we expect DM phenomenology to be directly related to the cosmological scale? One of the simplest evidences for the existence of DM comes from the galactic rotation curves. The objects in the halos of disc galaxies rotate around the galactic center at a higher velocity than it would be predicted by the visible amount of matter in the galaxies using Newtonian gravity. The surveys of this phenomenon over thousands of galaxies \parencite{McGaugh:2016leg} reveal a surprising set of regularities.

The discrepancy in the galactic rotation curves appears only at a large distance from the galactic center, while it disappears at the short distances. The transition between these two phases happens at a certain acceleration scale, measured universally as $a_0 = (1.20 \pm  0.02) \times 10^{-10} \, \mathrm{m}/\mathrm{s}^2$ with very little variation \parencite{Lelli:2017vgz}. Surprisingly, the existent variation does not correlate with any global or local galaxies' properties, hinting to a sort of \emph{Kepler's law} for galaxies, as phrased by McGaugh et al. Moreover, this value coincides in its order of magnitude with the cosmological scale,
\begin{equation}
    a_0 \approx c^2 \sqrt{\Lambda} \approx c H_0 \nonumber
    \;.
\end{equation}
One manifestation of this acceleration relation is observed for disk galaxies, where the asymptotic circular velocity $v_c$ is directly correlated with the total baryonic mass $M_b$ contained within the galaxy, 
\begin{equation}
    v_c^4 \sim M_b \nonumber
    \;.
\end{equation}
This relation is known as Baryonic Tully Fisher Relation. The proportionality constant is found to be $G a_0$, relating to the cosmological scale. Many other manifestations of the acceleration relation appears for other kinds of galaxies \parencite{Lelli:2017vgz}.

These relations describe the data empirically with no reference to the amount of DM in the galaxy. This implies that, at least in certain cases, the baryonic matter determines completely DM phenomenology with a single parameter, without the need for additional degrees of freedom. Explaining these observations poses a challenge for DM theories and provide the motivation to look for modifications to GR in the IR regime.

On the other hand, it is important to emphasize that there is overwhelming evidence for DM other than the galactic rotation curves. The most relevant ones come from Cosmology, in particular the role DM plays in structure formation as well as in the CMB power spectrum. Therefore, a reasonable approach seems to be to look for ways of unifying the success of such empirical laws described above with the explanatory dynamics DM provides in cosmological scales. A recent example of such an approach is provided by dark superfluids (see \cite{Berezhiani:2015bqa}).

Still, there is no fully-fledged theory that accounts for the whole spectrum of DM phenomenology, including the emergence of the acceleration scale and its seemingly coincidental value with $\Lambda$. However, there has been one phenomenological proposal that has been able to fit most of the astrophysical evidence for DM using just a single parameter, that is  Modified Newtonian Dynamics (MOND) \parencite{MOND-first}. We now consider what we can learn from it in order to help us establish what the $\Lambda$-theory should potentially recover in galactic scales. 

MOND makes the following proposal: If an object is accelerated above the critical acceleration $a_0$, $a \gg a_0$, it follows Newton's law $a = a_N(r) \equiv GM/r^2$; if it is accelerated below it, $a \ll a_0$, then it follows $a = \sqrt{a_0 a_N}$. This simple modification, despite its numerous problems, successfully predicts the galactic rotation curves with no additional matter and very small scattering, precisely as observed. 


If there is, indeed, an underlying gravity theory with MOND-like dynamics, then it is reasonable to expect the MOND rule $a = \sqrt{a_0 a_N}$ to follow from a deeper physical principle. One such candidate is the scale symmetry given by $t \rightarrow \lambda t$ and $r \rightarrow \lambda r$. Since the acceleration scales as $a \rightarrow \lambda^{-1} a$, this chooses a $1/r$ behavior for gravity. Hence, the MOND rule, unlike Newton's law, is scale-invariant in the low acceleration regime \parencite{Milgrom:2008cs}. It is fair to say that it is still not clear if this symmetry is incidental or fundamental, but if taken as fundamental then MOND-like dynamics may be revealing a symmetry to be pursued while constructing an extension to GR.

In fact, from that perspective, gravitational phenomena seems to have two different low-energy limiting regimes: i) one that preserves scale-invariance at low acceleration, ii) followed by another that breaks it for higher acceleration, recovering Newtonian gravity. The challenge is not only to build a satisfying mechanics that allows that sort of symmetry breaking, but also to be relativistic, and therefore extendable to cosmological scales.  


There have been many attempts\footnote{See e.g.~\parencite{Verlinde:2016toy} for a recent proposal.} to develop these ideas further and to construct a complete theory of modified gravity with MOND-like dynamics, so far with no definitive success. At the same time, direct observation attempts for DM have also failed, leaving its supposedly particle-like nature largely undetermined. While the efforts in both directions continue, our main point here is to emphasize that the puzzles appear to be tangling galactic and cosmological scales, and a consistent quantity, $\Lambda$, buried deeply in the IR, keeps haunting us as if defining an effective scale for which new physics should be expected. For now, we wait for the next chapters.

\section{The Cosmological Principle of Irreversibility}
Lastly, but not least, let us talk about the nature of time in Cosmology. This is possibly among the most involving topics both in Physics and Philosophy, and already has great representations in the context of Mythology. Maybe it is not a coincidence that one of the personifications of time in Greek mythology corresponded to a titan, \emph{Kronos}. And we are probably no \emph{Zeus} to defeat it. 

It is typically said that after the Theory of Relativity had been formulated, time and space became interchangeable. However, that is misleading: if this were true, the causal structure of spacetime would be lost altogether. 

Of course it remains true that the absolute notion of space and time is relative to each observer. But once a single observer is chosen, then there exists a clear notion of the flow of time; when the observer is changed, the flow of time may be different in comparison. And it happens to be the case that this flow is always pointing towards the future, regardless of the observer. To be more precise, given that the very notions of past and future are not easily translatable between different observers, for each and every observer the invariant notion about the flow of time is better encoded in the idea that their observed world evolves in an irreversible way. That is an empirical fact.


What about adiabatic reversible processes\footnote{Adiabatic processes happen without transfer of heat or mass between the environment and the system being considered. According to the first law of thermodynamics, that implies energy is transferred to the system only as work.}? They are non-existent, or more precisely, adiabaticity is only possible after a subsystem is considered. To say a process is adiabatic is just an idealization, usually quite satisfactory since the time scales of the system are much faster than the time scales of heat and/or mass being transferred between the system and its surroundings. Such an idealization allows us to have a first approximation of how the World works. The question for us is whether this is a reasonable assumption in Cosmology, since there is no such distinction between the system and its surroundings, given the Universe is all that exists. 

To tackle this question, first we note that the cosmological background in the context of the Standard Model emulated by a perfect fluid evolves following Friedmann equations, which imply the continuity equation, 

\begin{equation}
    \rho' + 3 H (\rho + p) = 0 \nonumber.
\end{equation}
This is the same as having a perfect fluid in a box under adiabatic expansion, thus it is reversible. Therefore, even though we observe the empirical fact that everything is walking towards future irreversibly whenever we pick up a local reference frame, somehow we miss this ingredient when coarse graining the Universe in order to account for its evolution. 

What is troublesome about a reversible adiabatic expansion in a cosmological background is that there is only one length scale available, defined by the Hubble parameter, $H^{-1}$, typically called Hubble radius. For most of the cosmological scenarios, $H^{-1}$ is linear in time. Thus our cosmic ruler can grow or shrink. When we have a fixed ruler with a single unit, like a meter stick, everything that is smaller than the ruler size is missed; that information is gone. Therefore, for a growing ruler, for instance, information is constantly being lost; more technically, ultraviolet modes, defined to be smaller than the Hubble radius, are being integrated out. And that, by the very definition of entropy, should result in an entropy increase of the system. 

Hence, it is crucial that the expansion of the Universe is described by laws that take into account that a cosmological background is constantly yielding entropy by missing out smaller modes than its characteristic scales. That is not the case for a perfect fluid in a cosmological background. 

Therefore, we propose \textbf{the Cosmological Principle of Irreversibility:} \emph{A Cosmological Theory should be inherently irreversible, and reversibility should be recovered only upon a truncation to a subsytem of the Universe.}

Since we expect the time-asymmetry to become manifest at the scales of the Universe, the Cosmological Principle of Irreversibility provides another guiding principle for the $\Lambda$-theory, which we argued as an IR extension of GR in the last section. Note that time-asymmetric dynamics have also been considered at the Planck scale and the most notable development in this direction can be found in \parencite{Cortes:2013uka} and subsequent work. Time asymmetric extensions of GR have also been studied in \parencite{Cortes:2015ola,Cortes:2016mfg}.


\section{Conclusions and Discussions}
The Standard Model of Cosmology, together with the Standard Model of Particle Physics, is among the best theories ever conceived. Both explain concisely a gigantic amount of data, and any possible attempt to extend these models will have to do an even better job. That makes the task quite challenging, but it remains necessary. 

As we have discussed, some problems keep haunting us despite our best efforts over many decades. We believe that we will need to rely on different fields of knowledge to help us with new insights on how to tackle them. 

Philosophy has played a major role in the development of physical theories in the past, and we believe it may still provide us with tools that may be lacking in our current approaches. It goes without saying that the philosophical ways of thinking allow us to have a clearer picture of the aforementioned problems, either because they highlight our assumptions and biases or because they provide a guide from first principles on how to construct theories. 

In this essay, we have attempted to propose, if not rephrase, some ideas concerning more extended versions of the Copernican Principle. This allowed us to consider the current evidence for MOND-like dynamics together with the evidence for the accelerated expansion of the late-time Universe to be hinting at something more fundamental about the structure of the Universe in the deep-infrared, possibly the presence of a new theory that deviates from GR, and is represented by the scales defined by the cosmological constant $\Lambda$.

Moreover, we laid down an argument for which one should expect that the Universe evolves irreversibly, so that a complete cosmological theory should be time asymmetric. We formulated this idea as the Cosmological Principle of Irreversibility. The arrow of time and the intrinsically time-asymmetric dynamics of the Universe should become manifest at the cosmological scale.

We have also discussed more abstractly the presence of hierarchies in our descriptions of the world, and elucidated the importance of avoiding any sort of infinite regress, leading us to introduce the Cosmological Heterarchical Principle. This should be seen as a guiding principle to tackle singularities in the very early universe. We have also shown that some elements of String Theory can be seen as a proxy for such a principle, thus avoiding the singularity altogether.

Our hope is that these philosophical principles and ideas that we underlined here will contribute to the development of novel theories of Nature.

\begin{center}
\Yingyang 
\end{center}

We did not just happen to be sitting by the fire. As with the Universe, we, humans, have also our own tale. Our story involves billions of years of evolution, which has shaped our bodies and minds to be fit to a very particular environment, embedded in very specific scales of space, time and energy. 

Through all the patterns that have come our way, some intuitions of the World have been formed and have been ingrained in us to degrees that are still challenging to be understood even when considering modern neuroscience. Two of them seem to be quite remarkable: cause and effect, and the finitude of things. Not surprisingly, those two intuitions have become precious ingredients to our understanding of the World, either in Physics or in any other field of study.

Cosmology, though, from its very beginning to its current state has been increasingly defiant in bending to our evolutionary prejudices. For when we consider the Universe as having any sort of beginning, our notions of cause and effect are directly challenged, while when we consider it to have any sort of eternal existence, then we cannot grasp its eternity, an infinity in time.

This may be seen as a pessimistic conclusion: it may be alluding to the fact that we are simply not up to the task, in the same way that we think that ants are not up to the task of understanding Quantum Mechanics or building cars; it may be hinting to the fact that some of our evolutionary prejudices are too ingrained in our approach to understand the Cosmos. Regardless of what this means to our physical theories of the moment, it is certainly a humbling lesson. It is a reminder of the human condition imprinted in our attempts to understand the world around us. And perhaps experiencing this lesson and the sense of wonder contained within it is more satisfying than having a final answer - at least for a while.

\section{Acknowledgements}
We are thankful to Lee Smolin and Zeguro Garrigues for many discussions, and Dan Petrescu, Renato Costa, Benjamin Bose and Megan Cowie for reading parts of the manuscript. The research at McGill is supported in part by funds from NSERC, from the Canada Research Chair program and from a John Templeton Foundation grant to the University of Western Ontario. GF is also thankful to University of Cape Town for hospitality during the period in which this work was written. This research was supported in part by Perimeter Institute for Theoretical Physics. Research at Perimeter Institute is supported by the Government of Canada through Industry Canada and by the Province of Ontario through the Ministry of Research and Innovation. This research was also partly supported by grants from NSERC, John Templeton Foundation, and FQXi.

\printbibliography

@article{Bernardo:2019pnq,
      author         = "Bernardo, Heliudson and Brandenberger, Robert and
                        Franzmann, Guilherme",
      title          = "{T-Dual Cosmological Solutions of Double Field Theory
                        II}",
      year           = "2019",
      eprint         = "1901.01209",
      archivePrefix  = "arXiv",
      primaryClass   = "hep-th",
      SLACcitation   = "%%CITATION = ARXIV:1901.01209;%%"
}

@article{Brandenberger:2018bdc,
      author         = "Brandenberger, Robert and Costa, Renato and Franzmann,
                        Guilherme and Weltman, Amanda",
      title          = "{T-dual cosmological solutions in double field theory}",
      journal        = "Phys. Rev.",
      volume         = "D99",
      year           = "2019",
      number         = "2",
      pages          = "023531",
      doi            = "10.1103/PhysRevD.99.023531",
      eprint         = "1809.03482",
      archivePrefix  = "arXiv",
      primaryClass   = "hep-th",
      SLACcitation   = "%%CITATION = ARXIV:1809.03482;%%"
}

@book{unger2015singular,
  title={The Singular Universe and the Reality of Time},
  author={Unger, R.M. and Smolin, L.},
  isbn={9781107074064},
  lccn={2014016833},
  url={https://books.google.co.za/books?id=IZFEBQAAQBAJ},
  year={2015},
  publisher={Cambridge University Press}
}

@InCollection{sep-cosmology,
	author       =	{Smeenk, Christopher and Ellis, George},
	title        =	{Philosophy of Cosmology},
	booktitle    =	{The Stanford Encyclopedia of Philosophy},
	editor       =	{Edward N. Zalta},
	note =	{\url{https://plato.stanford.edu/archives/win2017/entries/cosmology/}},
	year         =	{2017},
	edition      =	{Winter 2017},
	publisher    =	{Metaphysics Research Lab, Stanford University}
}

@InCollection{edgeEllis2017,
	author       =	{Ellis, George},
	title        =	{Physics on Edge},
	booktitle    =	{Inference: International Review of Science},
    volume         = "03, Issue 02",
	note =	{\url{https://inference-review.com/article/physics-on-edge}},
	year         =	{2017},
	edition      =	{},
	publisher    =	{}
}

@InCollection{sep-structural-realism,
	author       =	{Ladyman, James},
	title        =	{Structural Realism},
	booktitle    =	{The Stanford Encyclopedia of Philosophy},
	editor       =	{Edward N. Zalta},
	howpublished =	{\url{https://plato.stanford.edu/archives/win2016/entries/structural-realism/}},
	year         =	{2016},
	edition      =	{Winter 2016},
	publisher    =	{Metaphysics Research Lab, Stanford University}
}

@article{Smolin:1994vb,
      author         = "Smolin, Lee",
      title          = "{The Fate of black hole singularities and the parameters
                        of the standard models of particle physics and cosmology}",
      year           = "1994",
      eprint         = "gr-qc/9404011",
      archivePrefix  = "arXiv",
      primaryClass   = "gr-qc",
      reportNumber   = "CGPG-94-3-5",
      SLACcitation   = "%%CITATION = GR-QC/9404011;%%"
}

@incollection{Everett1973-EVETTO-2,
	title = {The Theory of the Universal Wavefunction},
	editor = {B. DeWitt and N. Graham},
	author = {Hugh Everett},
	year = {1973},
	booktitle = {The Many-Worlds Interpretation of Quantum Mechanics},
	publisher = {Princeton UP}
}

@inproceedings{Steinhardt:1982kg,
      author         = "Steinhardt, Paul Joseph",
      title          = "{NATURAL INFLATION}",
      booktitle      = "{Nuffield Workshop on the Very Early Universe Cambridge,
                        England, June 21-July 9, 1982}",
      year           = "1982",
      pages          = "251-266",
      reportNumber   = "UPR-0198T",
      SLACcitation   = "%%CITATION = UPR-0198T;%%"
}

@article{Vilenkin:1983xq,
      author         = "Vilenkin, Alexander",
      title          = "{The Birth of Inflationary Universes}",
      journal        = "Phys. Rev.",
      volume         = "D27",
      year           = "1983",
      pages          = "2848",
      doi            = "10.1103/PhysRevD.27.2848",
      reportNumber   = "TUTP-83-1",
      SLACcitation   = "%%CITATION = PHRVA,D27,2848;%%"
}

@article{Susskind:2003kw,
      author         = "Susskind, Leonard",
      title          = "{The Anthropic landscape of string theory}",
      year           = "2003",
      pages          = "247-266",
      eprint         = "hep-th/0302219",
      archivePrefix  = "arXiv",
      primaryClass   = "hep-th",
      SLACcitation   = "%%CITATION = HEP-TH/0302219;%%"
}

@article{Guth:1980zm,
      author         = "Guth, Alan H.",
      title          = "{The Inflationary Universe: A Possible Solution to the
                        Horizon and Flatness Problems}",
      journal        = "Phys. Rev.",
      volume         = "D23",
      year           = "1981",
      pages          = "347-356",
      doi            = "10.1103/PhysRevD.23.347",
      note           = "[Adv. Ser. Astrophys. Cosmol.3,139(1987)]",
      reportNumber   = "SLAC-PUB-2576",
      SLACcitation   = "%%CITATION = PHRVA,D23,347;%%"
}

@article{Linde:1981mu,
      author         = "Linde, Andrei D.",
      title          = "{A New Inflationary Universe Scenario: A Possible
                        Solution of the Horizon, Flatness, Homogeneity, Isotropy
                        and Primordial Monopole Problems}",
      booktitle      = "{QUANTUM COSMOLOGY}",
      journal        = "Phys. Lett.",
      volume         = "108B",
      year           = "1982",
      pages          = "389-393",
      doi            = "10.1016/0370-2693(82)91219-9",
      note           = "[Adv. Ser. Astrophys. Cosmol.3,149(1987)]",
      reportNumber   = "LEBEDEV-81-229",
      SLACcitation   = "%%CITATION = PHLTA,108B,389;%%"
}

@InCollection{sep-scientific-underdetermination,
	author       =	{Stanford, Kyle},
	title        =	{Underdetermination of Scientific Theory},
	booktitle    =	{The Stanford Encyclopedia of Philosophy},
	editor       =	{Edward N. Zalta},
	howpublished =	{\url{https://plato.stanford.edu/archives/win2017/entries/scientific-underdetermination/}},
	year         =	{2017},
	edition      =	{Winter 2017},
	publisher    =	{Metaphysics Research Lab, Stanford University}
}

@article{Ade:2015xua,
      author         = "Ade, P. A. R. and others",
      title          = "{Planck 2015 results. XIII. Cosmological parameters}",
      collaboration  = "Planck",
      journal        = "Astron. Astrophys.",
      volume         = "594",
      year           = "2016",
      pages          = "A13",
      doi            = "10.1051/0004-6361/201525830",
      eprint         = "1502.01589",
      archivePrefix  = "arXiv",
      primaryClass   = "astro-ph.CO",
      SLACcitation   = "%%CITATION = ARXIV:1502.01589;%%"
}

@book{Weinberg94,
  title={Dreams of a Final Theory},
  author={ Weinberg, S},
  lccn={},
  url={},
  year={1994},
  publisher={Vintage, New York}
}

@ARTICLE{2018FoPh...48..481R,
       author = {{Rovelli}, Carlo},
        title = "{Physics Needs Philosophy. Philosophy Needs Physics}",
      journal = {Foundations of Physics},
         year = "2018",
        month = may,
       volume = {48},
        pages = {481-491},
          doi = {10.1007/s10701-018-0167-y},
archivePrefix = {arXiv},
       eprint = {1805.10602},
 primaryClass = {physics.hist-ph},
       adsurl = {https://ui.adsabs.harvard.edu/\#abs/2018FoPh...48..481R}
}

@article{Price:1996,
author = {Price, Huw},
year = {1996},
month = {01},
pages = {},
title = {Time's Arrow \& Archimedes' Point: New Directions for the Physics of Time},
volume = {107},
journal = {The Philosophical Review},
doi = {10.2307/2998455}
}

@incollection{Penrose:1980ge,
      author         = "Penrose, R.",
      title          = "{SINGULARITIES AND TIME ASYMMETRY}",
      booktitle      = "General Relativity: An Einstein Centenary Survey",
      pages          = "581-638",
      year           = "1980",
      SLACcitation   = "%%CITATION = INSPIRE-159041;%%"
}

@article{Smolin:2005mq,
      author         = "Smolin, Lee",
      title          = "{The Case for background independence}",
      year           = "2005",
      pages          = "196-239",
      eprint         = "hep-th/0507235",
      archivePrefix  = "arXiv",
      primaryClass   = "hep-th",
      SLACcitation   = "%%CITATION = HEP-TH/0507235;%%"
}

@article{Esfeld2008-ESFMSR,
	year = {2008},
	volume = {160},
	pages = {27--46},
	journal = {Synthese},
	publisher = {Springer},
	number = {1},
	author = {Michael Esfeld and Vincent Lam},
	title = {Moderate Structural Realism About Space-Time}
}

@InCollection{sep-leibniz,
	author       =	{Look, Brandon C.},
	title        =	{Gottfried Wilhelm Leibniz},
	booktitle    =	{The Stanford Encyclopedia of Philosophy},
	editor       =	{Edward N. Zalta},
	howpublished =	{\url{https://plato.stanford.edu/archives/sum2017/entries/leibniz/}},
	year         =	{2017},
	edition      =	{Summer 2017},
	publisher    =	{Metaphysics Research Lab, Stanford University}
}

@article{AmelinoCamelia:2011bm,
      author         = "Amelino-Camelia, Giovanni and Freidel, Laurent and
                        Kowalski-Glikman, Jerzy and Smolin, Lee",
      title          = "{The principle of relative locality}",
      journal        = "Phys. Rev.",
      volume         = "D84",
      year           = "2011",
      pages          = "084010",
      doi            = "10.1103/PhysRevD.84.084010",
      eprint         = "1101.0931",
      archivePrefix  = "arXiv",
      primaryClass   = "hep-th",
      SLACcitation   = "%%CITATION = ARXIV:1101.0931;%%"
}

@InCollection{sep-infinite-regress,
	author       =	{Cameron, Ross},
	title        =	{Infinite Regress Arguments},
	booktitle    =	{The Stanford Encyclopedia of Philosophy},
	editor       =	{Edward N. Zalta},
	howpublished =	{\url{https://plato.stanford.edu/archives/fall2018/entries/infinite-regress/}},
	year         =	{2018},
	edition      =	{Fall 2018},
	publisher    =	{Metaphysics Research Lab, Stanford University}
}

@book{Hofstadter:1979:GEB:539932,
 author = {Hofstadter, Douglas R.},
 title = {Godel, Escher, Bach: An Eternal Golden Braid},
 year = {1979},
 isbn = {0465026850},
 publisher = {Basic Books, Inc.},
 address = {New York, NY, USA},
}

@article{Brandenberger:2018wbg,
      author         = "Brandenberger, Robert H.",
      title          = "{Beyond Standard Inflationary Cosmology}",
      year           = "2018",
      eprint         = "1809.04926",
      archivePrefix  = "arXiv",
      primaryClass   = "hep-th",
      SLACcitation   = "%%CITATION = ARXIV:1809.04926;%%"
}

@article{ArkaniHamed:2000tc,
      author         = "Arkani-Hamed, Nima and Hall, Lawrence J. and Kolda,
                        Christopher F. and Murayama, Hitoshi",
      title          = "{A New perspective on cosmic coincidence problems}",
      journal        = "Phys. Rev. Lett.",
      volume         = "85",
      year           = "2000",
      pages          = "4434-4437",
      doi            = "10.1103/PhysRevLett.85.4434",
      eprint         = "astro-ph/0005111",
      archivePrefix  = "arXiv",
      primaryClass   = "astro-ph",
      reportNumber   = "UCB-PTH-00-15",
      SLACcitation   = "%%CITATION = ASTRO-PH/0005111;%%"
}

@article{Smolin:2017kkb,
      author         = "Smolin, Lee",
      title          = "{MOND as a regime of quantum gravity}",
      journal        = "Phys. Rev.",
      volume         = "D96",
      year           = "2017",
      number         = "8",
      pages          = "083523",
      doi            = "10.1103/PhysRevD.96.083523",
      eprint         = "1704.00780",
      archivePrefix  = "arXiv",
      primaryClass   = "gr-qc",
      SLACcitation   = "%%CITATION = ARXIV:1704.00780;%%"
}

@book{Baumann:2014nda,
      author         = "Baumann, Daniel and McAllister, Liam",
      title          = "{Inflation and String Theory}",
      publisher      = "Cambridge University Press",
      year           = "2015",
      url            = "http://www.cambridge.org/mw/academic/subjects/physics/theoretical-physics-and-mathematical-physics/inflation-and-string-theory?format=HB",
      series         = "Cambridge Monographs on Mathematical Physics",
      doi            = "10.1017/CBO9781316105733",
      eprint         = "1404.2601",
      archivePrefix  = "arXiv",
      primaryClass   = "hep-th",
      ISBN           = "9781316237182",
      SLACcitation   = "%%CITATION = ARXIV:1404.2601;%%"
}

@article{Verlinde:2016toy,
      author         = "Verlinde, Erik P.",
      title          = "{Emergent Gravity and the Dark Universe}",
      journal        = "SciPost Phys.",
      volume         = "2",
      year           = "2017",
      number         = "3",
      pages          = "016",
      doi            = "10.21468/SciPostPhys.2.3.016",
      eprint         = "1611.02269",
      archivePrefix  = "arXiv",
      primaryClass   = "hep-th",
      SLACcitation   = "%%CITATION = ARXIV:1611.02269;%%"
}

@article{Milgrom:2008cs,
      author         = "Milgrom, Mordehai",
      title          = "{The MOND limit from space-time scale invariance}",
      journal        = "Astrophys. J.",
      volume         = "698",
      year           = "2009",
      pages          = "1630-1638",
      doi            = "10.1088/0004-637X/698/2/1630",
      eprint         = "0810.4065",
      archivePrefix  = "arXiv",
      primaryClass   = "astro-ph",
      SLACcitation   = "%%CITATION = ARXIV:0810.4065;%%"
}

@ARTICLE{MOND-first,
   author = "Milgrom, Mordehai",
    title = "{A modification of the Newtonian dynamics as a possible alternative to the hidden mass hypothesis}",
  journal = {apj},
     year = 1983,
    month = jul,
   volume = 270,
    pages = {365-370},
      doi = {10.1086/161130},
   adsurl = {http://adsabs.harvard.edu/abs/1983ApJ...270..365M}
}

@article{McGaugh:2016leg,
      author         = "McGaugh, Stacy and Lelli, Federico and Schombert, Jim",
      title          = "{Radial Acceleration Relation in Rotationally Supported
                        Galaxies}",
      journal        = "Phys. Rev. Lett.",
      volume         = "117",
      year           = "2016",
      number         = "20",
      pages          = "201101",
      doi            = "10.1103/PhysRevLett.117.201101",
      eprint         = "1609.05917",
      archivePrefix  = "arXiv",
      primaryClass   = "astro-ph.GA",
      SLACcitation   = "%%CITATION = ARXIV:1609.05917;%%"
}

@article{Cortes:2013uka,
      author         = "Cortês, Marina and Smolin, Lee",
      title          = "{The Universe as a Process of Unique Events}",
      journal        = "Phys. Rev.",
      volume         = "D90",
      year           = "2014",
      number         = "8",
      pages          = "084007",
      doi            = "10.1103/PhysRevD.90.084007",
      eprint         = "1307.6167",
      archivePrefix  = "arXiv",
      primaryClass   = "gr-qc",
      SLACcitation   = "%%CITATION = ARXIV:1307.6167;%%"
}

@article{Berezhiani:2015bqa,
      author         = "Berezhiani, Lasha and Khoury, Justin",
      title          = "{Theory of dark matter superfluidity}",
      journal        = "Phys. Rev.",
      volume         = "D92",
      year           = "2015",
      pages          = "103510",
      doi            = "10.1103/PhysRevD.92.103510",
      eprint         = "1507.01019",
      archivePrefix  = "arXiv",
      primaryClass   = "astro-ph.CO",
      SLACcitation   = "%%CITATION = ARXIV:1507.01019;%%"
}

@article{Cortes:2015ola,
      author         = "Cortes, Marina and Gomes, Henrique and Smolin, Lee",
      title          = "{Time asymmetric extensions of general relativity}",
      journal        = "Phys. Rev.",
      volume         = "D92",
      year           = "2015",
      number         = "4",
      pages          = "043502",
      doi            = "10.1103/PhysRevD.92.043502",
      eprint         = "1503.06085",
      archivePrefix  = "arXiv",
      primaryClass   = "gr-qc",
      SLACcitation   = "%%CITATION = ARXIV:1503.06085;%%"
}

@article{Cortes:2016mfg,
      author         = "Cortês, Marina and Liddle, Andrew R. and Smolin, Lee",
      title          = "{Cosmological signatures of time-asymmetric gravity}",
      journal        = "Phys. Rev.",
      volume         = "D94",
      year           = "2016",
      number         = "12",
      pages          = "123514",
      doi            = "10.1103/PhysRevD.94.123514",
      eprint         = "1606.01256",
      archivePrefix  = "arXiv",
      primaryClass   = "gr-qc",
      SLACcitation   = "%%CITATION = ARXIV:1606.01256;%%"
}

@Article{Hawking:1969sw,
  author       = {Hawking, S. W. and Penrose, R.},
  title        = {{The Singularities of gravitational collapse and cosmology}},
  journal      = {Proc. Roy. Soc. Lond.},
  year         = {1970},
  volume       = {A314},
  pages        = {529-548},
  doi          = {10.1098/rspa.1970.0021},
  slaccitation = {%%CITATION = PRSLA,A314,529;%%},
}

@Article{Borde:1993xh,
  author        = {Borde, Arvind and Vilenkin, Alexander},
  title         = {{Eternal inflation and the initial singularity}},
  journal       = {Phys. Rev. Lett.},
  year          = {1994},
  volume        = {72},
  pages         = {3305-3309},
  archiveprefix = {arXiv},
  doi           = {10.1103/PhysRevLett.72.3305},
  eprint        = {gr-qc/9312022},
  primaryclass  = {gr-qc},
  slaccitation  = {%%CITATION = GR-QC/9312022;%%},
}

@Article{Borde:2001nh,
  author        = {Borde, Arvind and Guth, Alan H. and Vilenkin, Alexander},
  title         = {{Inflationary space-times are incompletein past directions}},
  journal       = {Phys. Rev. Lett.},
  year          = {2003},
  volume        = {90},
  pages         = {151301},
  archiveprefix = {arXiv},
  doi           = {10.1103/PhysRevLett.90.151301},
  eprint        = {gr-qc/0110012},
  primaryclass  = {gr-qc},
  reportnumber  = {MIT-CTP-3183},
  slaccitation  = {%%CITATION = GR-QC/0110012;%%},
}

@Article{Brandenberger:1988aj,
  author       = {Brandenberger, Robert H. and Vafa, C.},
  title        = {{Superstrings in the Early Universe}},
  journal      = {Nucl. Phys.},
  year         = {1989},
  volume       = {B316},
  pages        = {391-410},
  doi          = {10.1016/0550-3213(89)90037-0},
  reportnumber = {HUTP-88-A035, BROWN-HET-673},
  slaccitation = {%%CITATION = NUPHA,B316,391;%%},
}

@Article{Huggett:2015vla,
  author        = {Huggett, Nick},
  title         = {{Target Space $\neq$ Space}},
  journal       = {Stud. Hist. Phil. Sci.},
  year          = {2016},
  volume        = {B59},
  pages         = {81-88},
  archiveprefix = {arXiv},
  booktitle     = {{Proceedings, Conference on The Philosophical Foundations of Dualities in Physics: Florence, Italy, September 15 - 16, 2014}},
  doi           = {10.1016/j.shpsb.2015.08.007},
  eprint        = {1509.06229},
  primaryclass  = {physics.hist-ph},
  slaccitation  = {%%CITATION = ARXIV:1509.06229;%%},
}

@Article{Lelli:2017vgz,
  author        = {Lelli, Federico and McGaugh, Stacy S. and Schombert, James M. and Pawlowski, Marcel S.},
  title         = {{One Law to Rule Them All: The Radial Acceleration Relation of Galaxies}},
  journal       = {Astrophys. J.},
  year          = {2017},
  volume        = {836},
  number        = {2},
  pages         = {152},
  archiveprefix = {arXiv},
  doi           = {10.3847/1538-4357/836/2/152},
  eprint        = {1610.08981},
  primaryclass  = {astro-ph.GA},
  slaccitation  = {%%CITATION = ARXIV:1610.08981;%%},
}

@article{Starobinsky:1980te,
      author         = "Starobinsky, Alexei A.",
      title          = "{A New Type of Isotropic Cosmological Models Without
                        Singularity}",
      journal        = "Phys. Lett.",
      volume         = "B91",
      year           = "1980",
      pages          = "99-102",
      doi            = "10.1016/0370-2693(80)90670-X",
      note           = "[,771(1980)]",
      SLACcitation   = "%%CITATION = PHLTA,B91,99;%%"
}

\end{document}